\documentstyle[12pt,aasms4]{article}

\newcommand{\be}{\begin{equation}}
\newcommand{\ee}{\end{equation}}

\lefthead{Wei \& Lu}
\righthead{The evolution of GRB remnants}

\begin{document}

\title{The evolution of GRB remnants}

\author{D.M. Wei}
\affil{Purple Mountain Observatory, Academia Sinica, Nanjing 210008, 
P.R.China} 
\and
\author{T. Lu}
\affil{Department of Astronomy, Nanjing University, Nanjing 210093, 
P.R. China\\
CCAST(World Laboratory) P.O.Box.8730, Beijing 100080, P.R.China\\
LCRHEA, IHEP, CAS, Beijing, P.R.China\\}
\authoraddr{pmoyl@pub.nj-online.nj.js.cn}

\begin{abstract}

The detection of the delayed emission in X-ray, optical and radio band,
i.e. the afterglow of $\gamma$-ray bursts (GRBs), suggestes that the sources
of GRBs are likely to be at cosmological distances. Here we explore
the interaction of a relativistic shell with a uniform interstellar
medium (ISM) and obtain the exact solution of the evolution of $\gamma$-ray
burst remnants, including the radiative losses. We show that in general
the evolution of bulk Lorentz factor $\gamma$ satisfies $\gamma \propto
t^{-\alpha_{t}}$ when $\gamma \gg 1$, here $\alpha_{t}$ is mainly in the
range $9/22 \sim 3/8$, the latter corresponds to adiabatic expansion.
So it is clear that adiabatic expansion is a good approximation even
when radiative loss is considered. However, in fact, $\alpha_{t}$ is
slightly larger than 3/8, which may have some effects on a detailed
data analysis. Synchrotron-self-absorption is also calculated and it
is demonstrated that the radio emission may become optically thin
during the afterglow. Our solution can also apply to the nonrelativistic
case ($\gamma \sim 1$), at that time the observed flux decrease more
rapidly than that in the relativistic case.

\end{abstract}


%\newpage

\section{Introduction}

Since the discovery of $\gamma$-ray bursts nearly 30 years ago, 
their origin and emission mechanisms remain mysterious, a major reason
is that their emissions other than X-ray and $\gamma$-ray have remained
invisible. This situation has been changed dramatically since the launch 
of Italian-Dutch satellite BeppoSax. The delayed emission at X-ray,
optical and radio wavelength has been detected from several GRBs due
to an accurate determination of their position (\cite{Costa97a,Costa97b})
The very long time afterglow (more than one month) strongly 
suggests that the sources of GRB are at cosmological distances.

It is a natural prediction of the cosmological fireball model that 
$\gamma$-ray bursts should have an afterglow in X-ray, optical and
radio bands (\cite{MR97,PR93,V97a}). 
After the main GRB event occurred, the fireball
continues to propagate into the ISM, and thus relativistic electrons 
may be continuously accelerated to produce the delayed radiation
on timescales of days to months. Several authors have discussed the 
radiation mechanisms of GRB afterglow and the agreement between the
fireball deceleration model and the measurements is good 
(\cite{W97a,W97b,WRM97}). However, these
authors simply assume that the fireball expansion is adiabatic. On the
other hand, Vietri (1997a; 1997b) has assumed that the fireball
expansion is highly radiative and the surrounding matter is nonuniform,
his results is also in agreement with the observations.

In this Letter we explore the interaction of a relativistic fireball 
with a uniform ISM. In section 2 we derive the exact solution of the
evolution of the bulk Lorentz factor $\gamma$ in general case,
without assuming expansion adiabatic or highly radiative.
Our solution can also extend to nonrelativistic case. In section 3 
we discuss the emission features for adiabatic expansion and
non-adiabatic expansion. The evolution of the synchrotron-self-absorption
frequency is also calculated. Finally we summarize our results and 
give some implications for future observations.

\section{The evolution of GRB remnant}

Whatever the sources of GRBs are, the observations suggest the following
scenario. Very large energy ($E \sim 10^{51}\, erg$ for cosmological
distances) is suddenly released ($T<100\,s$) in a small region of
space, so the initial energy density is so large that an opaque fireball
forms (\cite{P86,Good86,SP90}), then the
fireball expands outward relativistically. After an initial acceleration
phase, the fireball energy is converted to proton kinetic energy.
When the fireball is decelerated by the swept external matter, a strong
shock can be created and electrons may be accelerated to very high
energy. Then GRB is produced through synchrotron radiation or possible
inverse-Compton emission of these electrons. This occurs at decelerating
radius (\cite{MRP94})
\be
r_{d}=(\frac{2E}{\eta^{2}n_{1}m_{p}c^{2}})^{1/3}=5 \times 10^{16}
(\frac{E_{51}}{n_{1}})^{1/3}(\frac{\gamma_{0}}{100})^{-2/3}\;\; cm
\ee
where $E=10^{51}E_{51}$ is the total burst energy, $n_{1}$ is the density
of ISM, $\eta=E/M_{b}c^{2}$ represents the ratio of fireball energy to 
the initial rest mass energy of baryons, $M_{b}$ is the total mass of 
polluting baryons initially mixed with the fireball, and $\gamma_{0}$ is
the Lorentz factor of blast wave at radius $r_{d}$. It should be noted 
that at this time the shell thickness is similar to that of the heated
ISM region, so the heated ISM and shell carry similar energy 
(\cite{SP95,W97a}). Therefore it is easy to show that 
$\gamma_{0}=\eta/2$, and the ISM mass swept by the forward blast wave
at this time is $2/\eta$ of the shell's mass, i.e. $M_{0}/M_{b}=2/\eta$.

Sari \& Piran (1995) pointed out that the fireball energy 
dissipation may occur at two different places. They defined a radius
$r_{E}$ where the reverse shock becomes relativistic. If $r_{d}<r_{E}$,
a relativistic reverse shock can not be formed, and the shell loses its
energy to the heated ISM at $r_{d}$. Otherwise if $r_{d}>r_{E}$, then
most of the shell kinetic energy is converted into thermal energy once
the reverse shock crosses the shell (see also \cite{W97a}). This 
conclusion is valid only if the shell propagates with a constant width
$\Delta$. However, if the shell is expanding, then the shell width
$\Delta=r/\gamma^{2}$ will not be a constant. According to the definition
of Sari \& Piran (1995), the parameter $f(r)=n_{4}/n_{1}=\gamma l^{3}/
\eta r^{3}$, where $l=(\frac{E}{n_{1}m_{p}c^{2}})^{1/3}$ is the Sedov
length. Then when the reverse shock becomes relativistic, i.e. $\gamma^{2}/
f(r_{E})=1$, we can obtain
\be
r_{E}=(\frac{E}{\gamma\eta n_{1}m_{p}c^{2}})^{1/3}
\ee
comparing with equation (1), we find that $r_{E} \approx r_{d}$. So it
is clear that when the fireball expanding, the density ratio $f$ 
decrease with time. Initially $\gamma^{2}/f \ll 1$ and the reverse 
shock is Newtonian. When $r \sim r_{E}\,(r_{d})$ the fireball is
decelerated by ISM and at the same time the reverse shock become 
mildly relativistic.

After the GRB occurs, the fireball decelerates and a relativistic
blast wave continues to propagate into the ISM to produce the relativistic
electrons that may produce the delayed emission on time scales of
days to months. The light curves of the afterglow are controlled by
the deceleration of bulk Lorentz factor $\gamma$ and the intrinsic 
intensity $I_{\epsilon}'$. In previous researches some authors (\cite
{W97a,W97b,WRM97}) assume
that the fireball expansion is adiabatic, i.e. the radiative energy loss 
can be negligible, however some other people (\cite{V97a,V97b}) 
assume that the expansion is highly radiative, i.e. the internal energy 
of the system is instantaneously radiated out, it is obviously that
these are two extreme cases. In fact, the fireball should expand between 
these two cases, some fraction (not all) of the internal energy is radiated 
and lost from the system. So it is important to calculate the evolution
of bulk Lorentz factor $\gamma$ when including radiative energy loss.
We define $e_{s}$ to be the fraction of the internal energy 
emitted, then the total energy radiated per unit swept mass is given
by (\cite{BM76})
\be
dE=-\gamma(\gamma-1)e_{s}c^{2}dM
\ee
where $M=\frac{4}{3}\pi r^{3}n_{1}m_{p}$ is the swept ISM mass. 
For simplicity, we assume that the energy loss of the swept up
matter is small, then the total kinetic energy can be written as 
$E=(\gamma^{2}-1)Mc^{2}$. Thus we obtain
\be
\frac{d\gamma}{dM}=-\frac{(1+e_{s})\gamma^{2}-e_{s}\gamma-1}{2M\gamma}
\ee
which has the solution
\be
\frac{(\gamma-1)^{y}(\gamma+y^{-1})}{(\gamma_{0}-1)^{y}(\gamma_{0}+y^{-1})}
=(\frac{M}{M_{0}})^{-\frac{y(1+y)}{2}}
\ee
where $y=1+e_{s}$, $\gamma_{0}$ and $M_{0}$ are the initial values
of Lorentz factor and the mass of swept up matter at the radius
$r_{0}$ where the deceleration begins ($r_{0}\simeq r_{d}$).
When the expansion is highly relativistic ($\gamma \gg 1$), $\frac{\gamma}
{\gamma_{0}}=(\frac{M}{M_{0}})^{-\frac{y}{2}}=(\frac{r}{r_{0}})^{-
\frac{3y}{2}}$. If expansion is adiabatic ($e_{s}=0,\,y=1$), $\frac
{\gamma}{\gamma_{0}}=(\frac{r}{r_{0}})^{-3/2}$. Otherwise if expansion
is extremely radiative ($e_{s}=1,\,y=2$), then $\frac{\gamma}{\gamma_{0}}
=(\frac{r}{r_{0}})^{-3}$.
However, it should be pointed out that, in the above calculation we have taken 
the value of $e_{s}$ as a constant, this is appropriate for describing
the adiabatic and non-adiabatic limits ($e_{s}=0,\,1$), but for 
intermediate values the value of $e_{s}$ is almost certain not to be
a constant, and in fact probably depends on gas parameters which depend
on radius.

\section{the emission from GRB remnant}

The variation of bulk Lorentz factor $\gamma$ with observer time $t$
can be obtained by combining relation $dt=\frac{(1+z)dr}{2\gamma^{2}
\beta c}$ and equation (5), where $z$ is the cosmological redshift
of GRB source. For $\gamma \gg 1$ ($\beta \simeq 1$) it is given by
\be
\frac{\gamma}{\gamma_{0}}=(\frac{t}{t_{0}})^{-\alpha_{t}}
\ee
where $t_{0}=(1+z)r_{0}/2(1+3y)\gamma_{0}^{2}c$ is the typical 
duration of GRB event, $\alpha_{t}=3y/2(1+3y)$.
The fraction of the energy radiated can be estimated as $e_{s}=
e_{e}e_{syn}$, where $e_{e}$ represents the fraction of the energy that
is occupied by electrons, the typical Lorentz factor of electrons in
the comoving frame can usually be expressed as $\gamma_{e}=\xi_{e}\frac
{m_{p}}{m_{e}}(\gamma-1)$, so $e_{e}\simeq \frac{\xi_{e}}{1+\xi_{e}}$.
$\xi_{e}=1\,(e_{e}=\frac{1}{2})$ represents the energy equipartition
between electrons and protons. The synchrotron radiation efficiency 
can be expressed as $e_{syn}=\frac{t_{syn}^{-1}}{t_{syn}^{-1}+t_{ex}^{-1}}$
(we assume that synchrotron radiation is the main mechanism for GRB
afterglow), where $t_{syn}=6\pi m_{e}c/\sigma_{T}\gamma_{e}B^{2}$ and 
$t_{ex}=r/\gamma\beta c$ are the synchrotron cooling time and expansion
time in comoving frame respectively, $B=(\xi_{B}8\pi \gamma(\gamma-1)
n_{1}m_{p}c^{2})^{1/2}$ is the comoving frame magnetic field, $m_{e}
(m_{p})$ is the mass of electron (proton), $\sigma_{T}$ is Thomson 
cross section. Therefore we expect $e_{s}$ should lie in the range
$0 \sim 1/2$, and $y$ in the range $1 \sim 3/2$, so $\alpha_{t}$ should
between 3/8 (adiabatic expansion) to 9/22. Thus it is somewhat unexpected
that $\alpha_{t}$ lies in a very narrow range, for general case $\alpha_{t}$
is only slightly larger than 3/8 (adiabatic case), so it is reasonable
to think that adiabatic expansion is a good approximation and the
radiative energy loss could be treated as a small correction. Therefore
in the following we first consider the emission features at adiabatic
expansion.

{\em 3.1 adiabatic case}{\hspace{5mm}} From equation (5) we see that 
for adiabatic case ($y=1$) the bulk Lorentz factor $\gamma$ can be 
expressed as
\be
\gamma=(1+(\gamma_{0}^{2}-1)\frac{r_{0}^{3}}{r^{3}})^{1/2}
\ee
without any assumption. Here $r_{0}$ is the initial value of the 
deceleration phase, so $r_{0}=r_{d}$. From equation (1) $r_{d}^{3}=
2E/\eta^{2}n_{1}m_{p}c^{2}=E/2\gamma_{0}^{2}n_{1}m_{p}c^{2}=
l^{3}/2\gamma_{0}^{2}$, so $\gamma_{0}^{2}r_{d}^{3}=l^{3}/2$. Thus
for $\gamma_{0} \gg 1$ we have
\be
\gamma=(1+\frac{l^{3}}{2r^{3}})^{1/2}=(1+\frac{1}{x^{3}})^{1/2}
\ee
where $x=r/r_{c},\,r_{c}=l/2^{1/3}$, for $\gamma \gg 1$ from equation (6)
we obtain
\be
\gamma=278(1+z)^{3/8}E_{51}^{1/8}n_{1}^{-1/8}t^{-3/8}
\ee
This expression is valid only for relativistic expansion, it breaks down 
for $\gamma \simeq 1$. The time through which the blast wave remains 
relativistic is about a month.

If synchrotron radiation is the main mechanism to produce the GRB 
afterglow, we can estimate the time delay between the beginning of the
afterglow and the onset of the optical flash. The observed photon 
energy of synchrotron emission from typical electrons is ($\gamma \gg 1$)
\be
\epsilon_{m}=\frac{3m_{e}c^{2}B\gamma_{e}^{2}\gamma}{2(1+z)B_{c}}=5.2\times
10^{-3}\gamma^{4}n_{1}^{1/2}\xi_{e}^{2}\xi_{B}^{1/2}/(1+z) \;\;\;eV
\ee
where $B_{c}=4.413\times 10^{13}\,G$ is the critical magnetic field. Then
combining this equation with eq.(9), we have
\be
t_{m}\simeq 1.1\, (1+z)^{1/3}E_{51}^{1/3}(\frac{\epsilon_{m}}{1eV})^{-2/3}
\xi_{e}^{4/3}\xi_{B}^{1/3} \;\;(days)
\ee
Thus considering the uncerntain of the parameters, the value of $t_{m}$
(optical flash) could range from a few hours to about 3 days.

It can easily be seen from equation (10) that the typical photon energy
cannot possibly enter the radio region when $\gamma \gg 1$, so we
expect that until the fireball expansion is medium relativistic does
the synchrotron emission peak at radio band. It is interesting to note
that our solution (eq.(8)) can be extended to non-relativistic case. 
From eq.(8) the shell velocity $\beta=1/\sqrt{1+x^{3}}$, then the
relation between observer time and radial distance is $t=\int \frac{(1+z)
(1-\beta)dr}{\beta c}=\frac{(1+z)r_{c}}{c}\int (\sqrt{1+x^{3}}-1)dx$, 
so when $x \ll 1\,(\gamma \gg 1),\,r \propto t^{1/4}$, while when
$x \gg 1\,(\gamma \sim 1),\,r \propto t^{2/5}$.

The light curve of afterglow is also dependent on the intrinsic photon 
spectrum. We assume the comoving intensity is $I_{\epsilon}'\propto
\epsilon^{\alpha}$ for $\epsilon < \epsilon_{m}$ and $I_{\epsilon}'\propto
\epsilon^{\beta}$ for $\epsilon > \epsilon_{m}$. From equation (8), 
we can see, in the relativistic case, $\gamma_{e} \propto t^{-3/8},\,
B \propto \gamma,\, \epsilon_{m} \propto \gamma_{e}^{2}B\gamma \propto
t^{-3/2}, I_{\epsilon_{m}}' \propto n_{e}B\Delta r \propto t^{-1/8}$,
so the observed peak flux $F_{\epsilon_{m}} \propto t^{2}\gamma^{5}
I_{\epsilon_{m}}' \propto t^{0} \sim constant$ (\cite{MR97})
Before $\epsilon_{m}$ crosses the optical (radio) band, the
flux $F_{\epsilon} \propto F_{\epsilon_{m}}(\frac{\epsilon}{\epsilon_{m}})
^{\alpha} \propto t^{3\alpha/2}$, while after $\epsilon_{m}$ crosses 
the optical (radio) band, we have $F_{\epsilon} \propto F_{\epsilon_{m}}
(\frac{\epsilon}{\epsilon_{m}})^{\beta} \propto t^{3\beta/2}$. In the
non-relativistic case, $\gamma_{e} \propto t^{-6/5},\, B \propto t^{-3/5},
\, \epsilon_{m} \propto t^{-3},\, I_{\epsilon_{m}}' \propto t^{-1/5}$,
and $F_{\epsilon_{m}} \propto t^{3/5}$. So the observed optical (radio)
flux for $\epsilon < \epsilon_{m}$ is $F_{\epsilon} \propto F_{\epsilon_
{m}}(\frac{\epsilon}{\epsilon_{m}})^{\alpha} \propto t^{\frac{3}{5}
+3\alpha}$, for  
$\epsilon > \epsilon_{m}$ is $F_{\epsilon} \propto F_{\epsilon_
{m}}(\frac{\epsilon}{\epsilon_{m}})^{\beta} \propto t^{\frac{3}{5}
+3\beta}$. These features have also been discussed by Wijers, Rees 
\& M$\acute e$sz$\acute a$ros (1997). Fig.1 give an example of the
variation of observed flux with time t. We have taken $\alpha=0,\,
\beta=-1,\, E=5\times 10^{51}\,ergs$. The solid, dashed and dot-dashed
lines correspond to optical, x-ray and radio flux respectively. 
It can be seen from Fig.1 that the slope ($\epsilon > \epsilon_{m}$)
changes from -1.5 (relativistic case) to -2.4 (non-relativistic case)
gradually.

{\em 3.2 non-adiabatic case}{\hspace{5mm}} It can be seen from equation
(5) that for non-adiabatic case ($y >1$) we cannot obtain the explicit 
expression for the evolution of bulk Lorentz factor $\gamma$, only
when $\gamma \gg 1$ it can be written as the form of eq.(6). In order
to compare with the adiabatic case, we write $\alpha_{t}=3/8+\alpha'$,
where $\alpha'$ should be less than 3/88. Then we can write the
evolution of Lorentz factor as $\gamma =\gamma_{ad}(\frac{t}{t_{0}})^
{-\alpha'}$, $t_{0}\,(\sim 1-100\,s)$ is the typical duration of $\gamma$
ray flash, then we see that after one day $\gamma/\gamma_{ad}\sim 0.75$
for $t_{0}\sim 10\,s$, $\alpha'=3/88$, the exact value will depend on
the parameter $e_{s}$. In the meanwhile, from eq.(11) the time delay
between the GRB event and the onset of the optical flash will be
shorter by a factor $(t/t_{0})^{-\frac{8}{3}\alpha'}\sim 0.43$, i.e.
it could be range from a few hours to more than one day.

This will also affect the light curve of afterglow. Just as discussed
above, for relativistic case, $\gamma_{e} \propto t^{-(\frac{3}{8}+
\alpha')},\,
B \propto \gamma$, and $\epsilon_{m} \propto t^{-(\frac{3}{2}+4\alpha')}$.
It should be noted that, in the non-adiabatic case, the electron 
cooling time is shorter than the expansion time, and the effective 
width of the emission region is narrow than the width of swept up
matter, so
$I_{\epsilon_{m}}' \propto n_{e}Bct_{syn} \propto
t^{\frac{3}{8}+\alpha'}$ (\cite{MRW97}), 
and the 
observed peak flux $F_{\epsilon_{m}} \propto t^{\frac{1}{2}-4\alpha'}$ is not
a consant, it increases with time. Then the flux $F_{\epsilon} \propto
t^{(\frac{1}{2}+\frac{3}{2}\alpha)+4(\alpha-1)\alpha'}$ for 
$\epsilon < \epsilon_{m}$ and
$F_{\epsilon}\propto t^{(\frac{1}{2}+\frac{3}{2}\beta)+4(\beta-1)\alpha'}$
for $\epsilon > \epsilon_{m}$. Therefore we see that for $\epsilon <
\epsilon_{m}$ the flux $F_{\epsilon}$ may either increase or decrease 
with time, which depends on the values of $\alpha$ and $\alpha'$. 

For non-relativistic case the variation of $\gamma$ with $t$ is more
complicated. However, it should be pointed out that radiative energy
loss may be important only in early time. The ratio of synchrotron 
cooling time to the expansion time in the comoving frame is
\be
\frac{t_{syn}}{t_{ex}}=7\times 10^{6}\gamma_{0}^{-4}\xi_{e}^{-1}
\xi_{B}^{-1}n_{1}^{-1}t_{0}^{-1}(\frac{t}{t_{0}})^{4\alpha_{t}-1}
\ee
Taking $\alpha_{t}=9/22$, for $t_{syn}/
t_{ex}=1$ we obtain $t_{e}\simeq 0.4 \xi_{e}^{11/7}\xi_{B}^{11/7}
E_{51}^{6/7}n_{1}^{5/7}(\frac{\gamma_{0}}{100})^{-4/7}$ day. Therefore 
we expect that the synchrotron cooling time to be longer than the
dynamical time for long delay afterglow (usually $t_{e} <1$ day).
When $t_{syn} > t_{ex}$, the synchrotron radiation efficiency $e_{syn}$
decreases rapidly and so $e_{s} \ll 1$, at that time the adiabatic
expansion is a very good approximation.

{\em 3.3 synchrotron-self-absorption}{\hspace{5mm}} We have shown that
the typical photon energy $\epsilon_{m}$ might be impossible to
enter the radio region when $\gamma \gg 1$, so the detected radio flash
within a few days after GRB (\cite{Frail97}) may be possible due
to the source becoming optically thin. As Waxman (1997b) and Vietri (1997b) 
have pointed out, the synchrotron-self-absorption optical depth for
photon energy $\epsilon < \epsilon_{m}$ may scale as $\tau=\tau_{m}
(\frac{\epsilon}{\epsilon_{m}})^{-2}$ due to the presence of a low
energy electron population. The optical depth $\tau_{m}$ can be estimated
as $\tau_{m}\simeq 1.2\times 10^{-13}n_{1}^{1/2}\xi_{e}^{-5}\xi_{B}^{-1/2}
\gamma^{3/2}(\gamma-1)^{-11/2}\beta t$, then for adiabatic approximation,
the absorption frequency for $\gamma \gg 1$ is $\nu_{ab}\simeq 2n_{1}^
{1/2}\xi_{e}^{-1/2}\xi_{B}^{1/4}E_{51}^{1/4}t_{day}^{-1/4}\,GHz$, while
for $\gamma \sim 1$, $\nu_{ab} \simeq 0.2n_{1}^{3/4}\xi_{e}^{-1/2}
\xi_{B}^{1/4}t_{day}^{1/2}\, GHz$. So the absorption frequency first
decreases with time as $\nu_{ab} \propto t^{-1/4}$, while when the
expansion is non-relativistic, $\nu_{ab}\propto t^{1/2}$. When photon
energy $\epsilon > \epsilon_{m}$, $\tau \propto \epsilon^{-(p+4)/2}$,
where $p$ is the index of electron distribution, then $\nu_{ab}=
\nu_{m}\tau_{m}^{2/(p+4)}$. Taking $p=3$, then $\nu_{ab}=317n_{1}^{2/7}
\xi_{e}^{4/7}\xi_{B}^{5/14}E_{51}^{5/14}t_{day}^{-11/14}\, GHz$ for
$\gamma \gg 1$, and $\nu_{ab}=1556n_{1}^{3/14}\xi_{e}^{4/7}\xi_{B}^{5/14}
E_{51}^{3/7}t_{day}^{-1}\, GHz$ for $\gamma \sim 1$. The observed 
flux may consists of several components. If $\nu_{ab} <\nu_{m}$, then
$F_{\nu}=F_{\nu_{m}}(\frac{\nu_{ab}}{\nu_{m}})^{\alpha}(\frac{\nu}{\nu_{ab}})
^{\alpha+2}$ for $\nu <\nu_{ab}$, $F_{\nu}=F_{\nu_{m}}(\frac{\nu}{\nu_{m}})^
{\alpha}$ for $\nu_{ab} < \nu <\nu_{m}$, and $F_{\nu}=F_{\nu_{m}}(\frac{\nu}
{\nu_{m}})^{\beta}$ for $\nu >\nu_{m}$. If $\nu_{ab} > \nu_{m}$, then
$F_{\nu}=F_{\nu_{m}}(\frac{\nu_{ab}}{\nu_{m}})^{\beta}(\frac{\nu}{\nu_{ab}})
^{5/2}$ for $\nu <\nu_{ab}$, and $F_{\nu}=F_{\nu_{m}}(\frac{\nu}{\nu_{m}})^
{\beta}$ for $\nu > \nu_{ab}$. These components can be seen in our 
calculated radio flux in Fig.1.

\section{Discussion and conclusion}

The detection of $\gamma$-ray burst in the optical and radio bands has
greatly furthered our understanding of the objects. Some people have
discussed the afterglow emission using the adiabatic expansion or
highly radiative expansion. Here we have investigated the evolution
of the fireball Lorentz factor $\gamma$ when it interacts with the ISM.
We find that even though including the radiative energy loss, the
difference of the evolution of the $\gamma$ between our results and that 
in adiabatic case is small. So adiabatic expansion can be treated 
as a good approximation.

From eq.(6) we can also obtain the evolution of the total system energy 
(when $\gamma \gg 1$) $E/E_{0}=(t/t_{0})^{-3e_{e}/(4+3e_{s})}$, which
is different from the results of Sari (1997), who obtained $E \propto
t^{-17e_{s}/16}$. This is because Sari had used the self-similar solution
to obtain his results, while our results are obtained by solving 
the energy loss equation without any assumption. Our results show
that the decrease of total energy is slower than that predicted
by Sari (1997).

We have shown that the energy loss due to radiation may be important 
only in early time of afterglow (usually $t<1$ day), since the 
synchrotron cooling time may be much longer than the expansion 
time for long delay emission, so we argue that the detection of 
afterglow within short time after the GRB event is very important,
since it may provide more information about the GRB sources, the
radiation processes, and the surrounding matter features, etc..

The afterglow has been detected for more than one month, the interesting
question is how long can the afterglow be sustained? From Fig.1 we
see that the optical flux decays with time as a power law $F \propto
t^{-n}$, with the index $n$ increases gradually from $\frac{3}{2}\beta$
to $(\frac{3}{5}+3\beta)$. Therefore we argue that the optical flux 
should be detected for longer time without suddenly cutoff. In
addition it is also possible to observe radio emission for very long time.

The detection of very long afterglow strongly suggests that the sources
of GRBs are at cosmological distances, which mean the total energy
of GRB event should be about $10^{51-52}\,ergs$. Now the most fundamental
problem, the ultimate energy source and the physical processes leading
to the fireball formation has not yet  been solved. Recently the 
popular model--coalescing of two neutron stars-- seems to be difficult
to account for the GRBs (\cite{RJTS97}), and now a new way, very
strong magnetic field ($B\sim 10^{15}\,G$) combined with rotation,
has been suggested to power the fireball (\cite{P97}). We expect
the detection of GRB afterglow can provide information about the
origin of GRBs.

\acknowledgements{We thank the referee for several important 
comments which improved this paper. This work was supported by
National Natural Science Foundation and the National Climbing 
Programme on Fundamental Researches of China.}

\newpage

\figcaption[after.ps]
{The evolution of observed flux with time $t$. We have taken $\alpha
=0,\,\beta=-1,\,E=5\times 10^{51}\,ergs$. The solid, dashed and dot-dashed
lines correspond to optical, X-ray and radio flux respectively.
\label{Fig.1}}

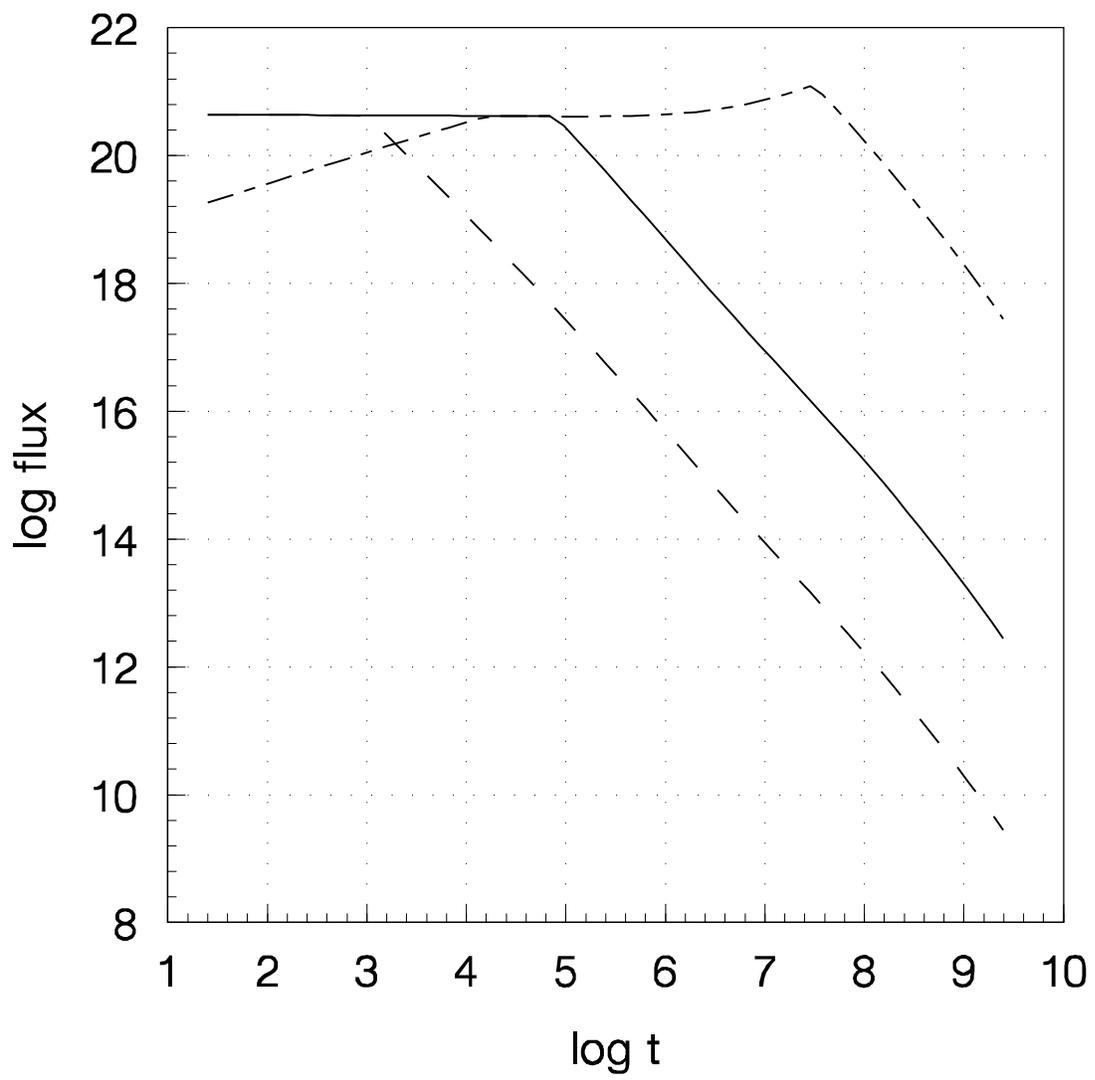
\end{document}